\documentclass[pra,superscriptaddress,twocolumn]{revtex4-1}
\usepackage{amssymb}
\usepackage{amsmath}
\usepackage{graphicx}
\usepackage[english]{babel}
\usepackage[T1]{fontenc}
\usepackage{float}
\usepackage{xcolor}
\newcommand{\be}{\begin{equation}}
\newcommand{\ee}{\end{equation}}

\graphicspath{{../../figuresN/}}
\begin{document}

\title{Nonlinear optical lattices with a void impurity}

\author{Cristian Mej\'{\i}a-Cort\'{e}s}
\email{ccmejia@googlemail.com} 
\author{J. C. Cardona}
\affiliation{Programa de F\'{\i}sica, Facultad de Ciencias B\'{a}sicas,
Universidad del Atl\'{a}ntico, Puerto Colombia 081007, Colombia}
\author{Andrey A. Sukhorukov}
\affiliation{Nonlinear Physics Centre, Research School of Physics and
Engineering, Australian National University, Canberra, ACT 2601, Australia}
\author{Mario I. Molina}
\affiliation{Departamento de F\'{\i}sica, Facultad de Ciencias, Universidad de
Chile, Casilla 653, Santiago, Chile}

\date{\today }

\begin{abstract} 
    We examine a one-dimensional nonlinear (Kerr) waveguide array which
    contains a single `void' waveguide where the nonlinearity is identically zero.
    We uncover a new family of nonlinear localized modes centered at
    or near the void, and their stability properties. Unlike a usual impurity
    problem, here the void acts like a repulsive impurity causing the center of the simplest mode to lie to the side of the void's position. We also
    compute the stability of extended nonlinear modes showing significant
    differences from the usual homogeneous nonlinear array. The
    transmission of a nonlinear pulse across the void shows three main regimes,
    transmission, reflection and trapping at the void's position, and we identify a
    critical momentum for the pulse below (above) where the pulse gets reflected
    (transmitted), or trapped indefinitely at the void's position. For relatively    wide pulses, we observe a steep increase from a regime of no transmission to a regime of high    transmission, as the amplitude of the soliton increases beyond a critical
    wavevector value.  Finally, we consider the transmission of an extended
    nonlinear wave across the void impurity numerically, finding a rather
    complex structure of the transmission as a function of wavevector, with the
    creation of more and more transmission gaps as nonlinearity increases. The
    overall transmittance decreases and disappears eventually, where the
    boundaries separating passing from non-passing regions are complex and
    fractal-like.
\end{abstract}

\maketitle

\section{Introduction}

The problem of the effect of impurities in otherwise periodic systems, is an old
topic. Point impurities or extended defects are known to break the shift translational invariance of linear discrete, periodic systems, leading to the formation of localized
modes, wave reflection and resonant scattering where energy or probability is
localized in the vicinity of the impurity~\cite{defect1a,defect1b, defect2}.
Perhaps the oldest example of this is a site or coupling defect in a
tight-binding system~\cite{TB1, TB2}. In this case, one finds  a mode that
decays exponentially away from the impurity site, which exists for any amount of
impurity strength and whose energy lies outside the band.  The presence of
judiciously placed defects make possible interesting resonance phenomena, such
as Fano resonances (FR), where there is total reflection of plane waves through
the impurity region, in an otherwise periodic potential. In a typical FR system,
the wave propagation in the presence of a periodic scattering potential is
characterized by open and closed channels. The open channel guides the
propagating waves as long as the eigenfrequencies of closed channels do not
match the spectrum of linear waves.  Total reflection of waves in the open
channel occurs when a localized state originating from one of the closed
channels resonates with the open channel spectrum~\cite{fano1,fano2}. 
Another
type of defect is provided by the junction between two periodic optical arrays,
or networks~\cite{christodoulides}. This type of defect is more of a topological
nature. Well-known systems for realizing these network junctions are photonic
crystal circuits and microring resonator structures with more than two coupled
channel waveguides. It has been proven ~\cite{junction} that for a junction with
more that two branches, there are two defect states localized at the junction.
The transmission of plane waves is always smaller than one, irrespective of the
number of branches. A kind of topologically-protected localized modes 
in periodic dielectric structures modulated by a domain wall has been examined in  \cite{topological}. Another context where isolated impurities can play a role is
in magnetic metamaterials, modeled as periodic arrays of split-ring resonators
(SRRs). Simple defects for a SRR include the capacitive defect, created when the
capacitance of one of the rings is changed by either altering the slit width, or
by inserting a dielectric in the slit. Another simple defect is the
inductive defect, caused by altering the distance between a single ring and its
closest neighbors~\cite{SRR}.  In all these cases it is possible to show that
magnetic energy can be trapped at impurity positions.

In the case of a single cubic nonlinear impurity embedded in a linear
lattice, it is possible to extend the formalism of lattice Green
functions~\cite{economou} to obtain exact expressions for the bound state energy
and for the transmission coefficient across the impurity. In this case, there is
a minimum nonlinearity strength required to create a localized mode at the
impurity site. This mode originates from the mode at the edge of the band that
detaches into the gap under the action of nonlinearity~\cite{PRB}. For
nonlinearity exponents higher than three, it is possible to find more than one
bound state, although not all of them are stable~\cite{generalized}. For the case
of a single cubic impurity in a square lattice of arbitrary nonlinearity
exponent, it was shown that two bound states (localized at the impurity) exist
above a certain exponent-dependent critical nonlinearity strength. The
localization length of the lower (higher) energy bound state increases
(decreases) with nonlinearity strength~\cite{square}. When the nonlinearity of
the impurity is of the saturable form, one finds that an impurity bound state
located at the impurity site is always possible, independent of the impurity
strength, while for the surface saturable impurity case, a minimum nonlinearity
strength is necessary. The transmission coefficient of plane waves across the impurity is always smaller than in the case of a linear impurity, and shows no resonances~\cite{PRE}.

In this work we consider a case complementary to the usual one~\cite{PRB},
consisting of a {\em void} waveguide embedded in an otherwise nonlinear (Kerr)
waveguide array. In other words, the array is nonlinear at all sites except at a
given site  where the nonlinearity of the waveguide is identically zero. Since
the background is fully nonlinear it is no longer possible to use the usual
Green function formalism since this formalism depends crucially on the linearity
of the system, and on the orthogonality and completeness of the base functions.
Therefore, most of the treatment will have to rely on numerics. 

As we will show, there is a novel family of localized modes in the vicinity of
the void impurity.  The extended nonlinear modes show substantially different
stability properties than for the homogeneous case.  Also, the transmission of a
nonlinear pulse across the impurity shows a distinct transition from passing to
non-passing regime as a function of the soliton amplitude, while the
transmission of plane waves across the void, shows a complex boundary between
passing and non-passing regimes, in wavevector-nonlinearity  parameter space.

\section{The model}
For definiteness, we will work in an optics context, although the methods can be
applied to other systems as well e.g., condensed matter \cite{condensed1,condensed2,condensed3,condensed4,condensed5,condensed6} 
for which the
discrete nonlinear Schr\"{o}dinger (DNLS) equation applies (see below).  
An optical context is ideal for the demonstration of basic
phenomenology concerning propagation of generic excitations on a periodic
discrete medium, the effect of nonlinearity and disorder, impurities and other
phenomena that are difficult to isolate and study independently due to the presence
of other degrees of freedom, many-body effects, etc. In fact, it was in an
optical setting that famous Anderson localization was first observed in a clean
way~\cite{anderson1,anderson2}.  

Thus, let us consider the propagation of light along a one-dimensional nonlinear
(Kerr) waveguide array with unit periodicity which contains a single {\em void}
impurity at $n=n_{c}$. This void  consists on a purely linear waveguide.
In the framework of the coupled-mode theory, the evolution equations for the
electric field amplitudes $E_{n}(z)$ can be written in the form
\be i {d
E_{n}\over{d z}} + \epsilon E_{n} + \sum_{m} V_{nm} E_{m} +
(1-\delta_{n,n_{c}})\ \chi\  |E_{n}|^2 E_{n}=0,\label{eq:1} 
\ee
known as the
discrete nonlinear Schr\"{o}dinger (DNLS) equation, where $E_{n}(z)$ is the
electric field [in units of $(\mbox{Watt})^{1/2}$] on the $nth$ waveguide at
distance $z$ (in meters), $V_{n m}$ is the evanescent coupling (in units of 1/m)
between guide $n$ and $m$, $\epsilon$ is the linear index of refraction and
$\chi$ [in units of $1/(\mbox{Watt}\times m)$] is the nonlinear coefficient,
defined as $\chi=\omega_{0} n_{2}/c\ \mbox{A}_{\mbox{eff}}$, where $\omega_{0}$
is the angular frequency of light, $n_{2}$ is the nonlinear refractive index
of the guide, and $\mbox{A}_{\mbox{eff}}$ is the effective area of the linear
modes. Typical values for these parameters are: $z_{max}=5\times 10^{-3} m, V_{n m}=O(10) m^{-1},
\omega_{0}=2.4 \times 10^{15}, A_{eff}\sim 60\times 10^{-12}\ m^{2}, n_{2}=1.35 \times 10^{-20} m^{2}/W$, which implies $\chi\approx 1.8\times 10^{-3} (m W)^{-1}$. In the strong nonlinear regime, the optical power is about\ $10^{6} W$ which implies electric fields of the order of $10^{3} W^{1/2}$. 
Hereafter, we will take $\epsilon=0$, which only has the effect of
redefining the phase of $E_{n}$. Each waveguide can be fabricated by means of
the femtosecond direct writing technique~\cite{femto}, where the effective
nonlinearity of the guide is controlled by the laser power and the writing
speed. 

Equation (\ref{eq:1}) has two conserved quantities, the power
\be 
P=\sum_{n}|E_{n}(z)|^2, 
\ee
and the Hamiltonian 
\be 
H = -\sum_{m\neq n} V_{n m}
E_{n}(z) E_{m}^{*}(z) - \sum_{n} \frac{\chi}{2} (1-\delta_{n,n_{c}}) |E_{n}(z)|^4.
\ee
They are useful to monitor the accuracy of our numerical computations.

\section{Localized modes}
\label{localized modes}
Let us consider stationary states $E_{n}(z) = \exp(i \lambda z) C_{n}$, where the amplitudes 
$C_{n}$ satisfy
\be
-\lambda C_{n} + \sum_{m} V_{n m } C_{m} + (1-\delta_{n,n_{c}})\ \chi\ |C_{n}|^2 C_{n}=0.
\label{eq:2}
\ee
Since the waveguide array is intrinsically nonlinear, it is no longer possible
to use the technique of lattice Green functions to obtain the localized mode.
Instead, we will resort to the numerical method of Newton-Raphson, to solve the
system of nonlinear equations (\ref{eq:2}), and find the different classes of
modes that reside in the vicinity of the void, $n=n_{c}$.  For simplicity, we will consider coupling to nearest-neighbor guides only and set $V_{n, n+1}=V_{n,n-1}\equiv V>0$ and $\chi\geq 0$, conditions that ensure $\lambda>2 V$.  In addition
to determining the shape of the modes, we will also compute their linear
stability. The stability analysis starts from a perturbed stationary solution $E_{n}(t)=(C_{n}+\delta_{n}(z)) \exp(i \lambda z)$ which is introduced in the evolution equation (\ref{eq:1}), followed by an expansion in powers of the perturbation $\delta_{n}(z)$, keeping only linear terms in $\delta_{n}(z)$. This results in a linear equation for the perturbation 
$\delta_{n}$.  Writing $\delta_{n}=\xi_{n}+ i \eta_{n}$ yields something of the form
\[
{d\over{dt}} \left(
\begin{array}{c}
\xi_{n}\\
\eta_{n}\\
\end{array}
\right)  = \cal{M} 
\left(
\begin{array}{c}
\xi_{n}\\
\eta_{n}\\
\end{array}
\right) 
\]
with 
\be
\cal{M} = \begin{pmatrix} 
\mbox{\bf 0} & \mbox{\bf A} \\
\mbox{\bf B} & \mbox{\bf 0} 
\end{pmatrix} 
\ee
where 
$\mbox{\bf A}$ and $\mbox{\bf B}$ are $N\times N$ matrices given by
\be
\mbox{\bf A}_{n,m} = V(\delta_{n,m+1}+ \delta_{n,m-1})+ (2 \chi_{n}|u_{n}|^2+ \chi_{n} u_{n}^2-\lambda)\delta_{n,m}
\ee
\be
\mbox{\bf B}_{n,m} = V(\delta_{n,m+1}+ \delta_{n,m-1})+ (2 \chi_{n}|u_{n}|^2- \chi_{n} u_{n}^2-\lambda)\delta_{n,m}
\ee
and $\chi_{n}=\chi(1-\delta_{n,n_{c}})$.
The stability requirement is equivalent to the matrix $\cal{M}$ having all of its eigenvalues lying on the imaginary axis.

Results from this procedure are shown in Fig. 1 for $n_{c}\sim
N/2$ to minimize boundary effects. The mode (b)
shows a simple maximum at $n=n_{c}-1$, instead of $n=n_{c}$, which is the case
with usual non-void impurities. The reason is probably due to the fact that at $n_c$, nonlinearity drops to zero thus creating a sort of boundary between the nonlinear and the  void region. This abrupt boundary creates a surface
mode that decays towards the `void' and increases in the opposite direction \cite{OPN}. The region to the right of the void (which is entirely nonlinear) has almost no effect on the mode, since in that region the effective nonlinearity (nonlinearity $\times$   amplitude), is very small due to the small tail. On the other hand, by employing a symmetric seed around the void, our iterative scheme converges to mode (c), which looks like two
two type-(b) modes coexisting in the immediate vicinity of the void. However, this mode is unstable in its entire domain. 
It should be noted that, for the class of symmetric modes around the 
void, it can be shown that the amplitude at the void has to be a local minimum. The proof is simple:
Let us set $n_{c}=0$ for simplicity, then the stationary equation at the void reads
$-\lambda C_{0} + V (C_{1}+C_{-1})=0$. For a symmetric mode $C_{1}=C_{-1}$ implying
$C_{1}/C_{0}= \lambda/2 V$. Now,  since the localized mode is always outside the band, $\lambda>2 V$. Thus, $C_{0}/C_{1}<1$.
\begin{figure}[t]
  \includegraphics[width=\linewidth]{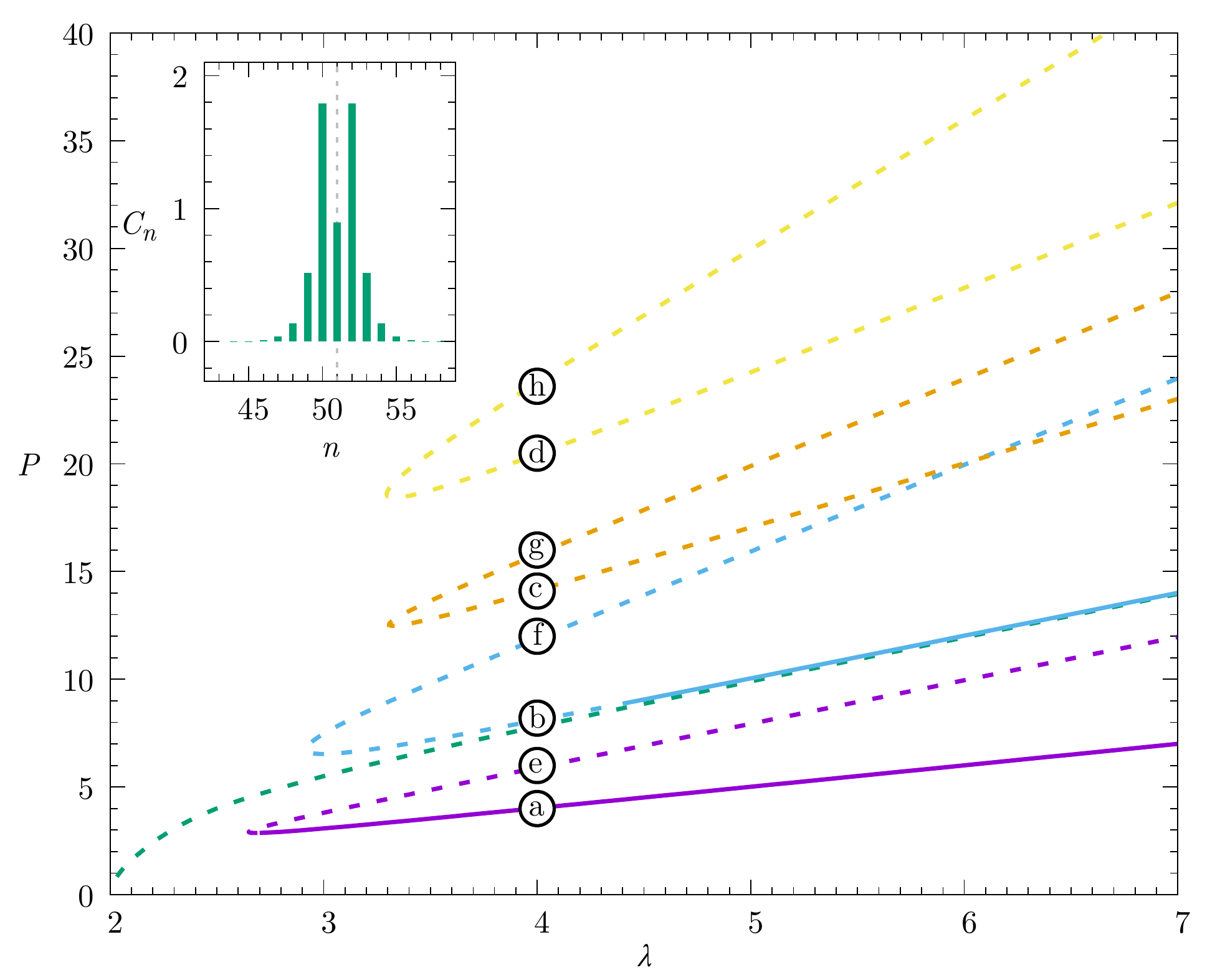}
  \includegraphics[width=\linewidth]{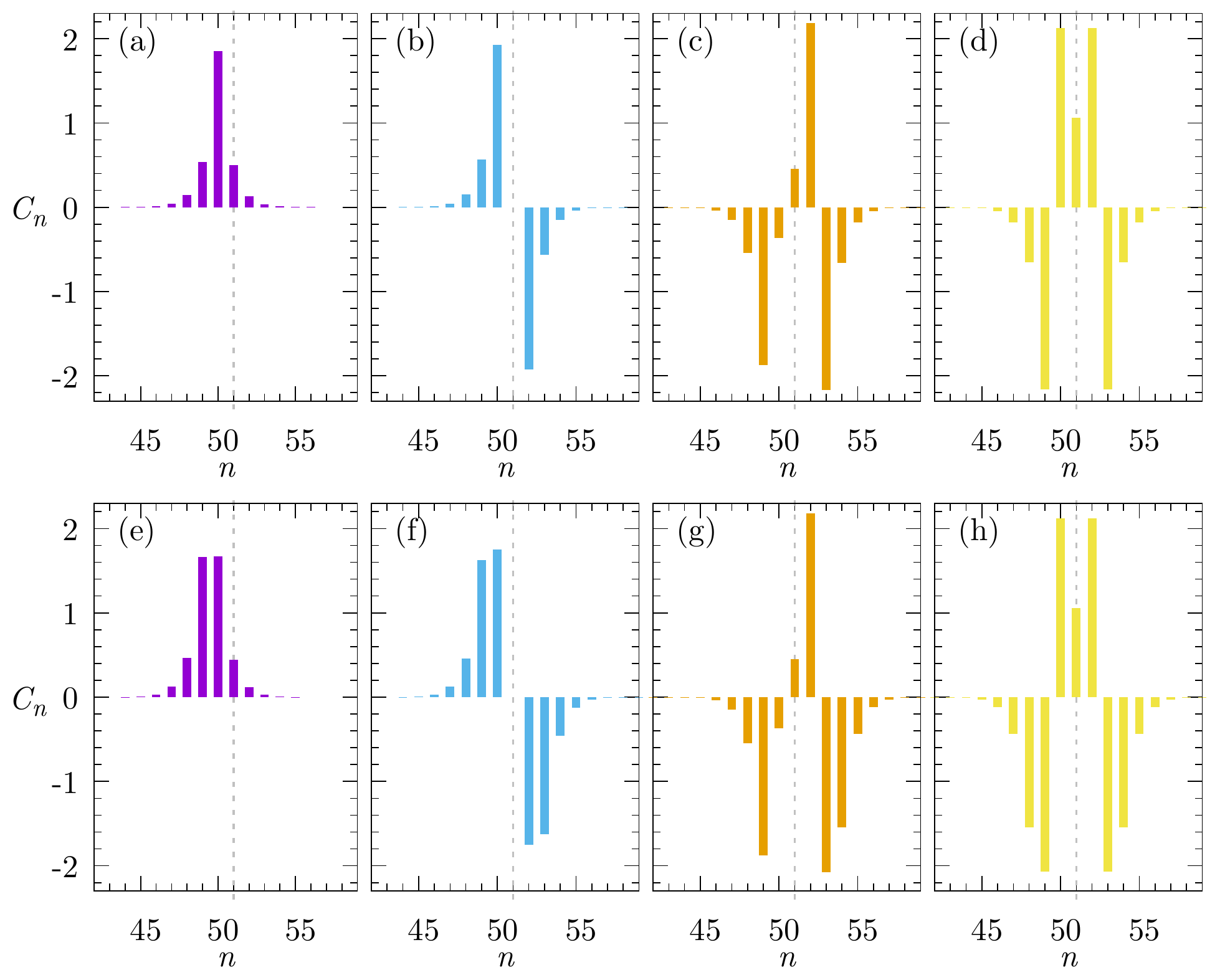}
  \caption{Top: Power vs eigenvalue curves for some families of low-lying modes, in
    the vicinity of the nonlinear void. The continuous (dashed) curves denote
    stable (unstable) regimes. The inset shows the profile of the mode whose power curve reaches all the way down to the band. Bottom: Spatial profiles of
    the  low-lying modes. Notice that none of the maxima falls on $n=n_{c}=51$.}
  \label{fig1}
\end{figure}
On the other hand, if we start from an anti
symmetric profile around the void, a new mode (d) appears,  exhibiting amplitude inversion. We call it a `twisted' mode~\cite{twisted1,twisted2}, where the amplitude vanishes at $n=n_{c}$. Unlike mode (b), the twisted mode  possesses a stability domain for $\lambda\gtrsim 4$. Since the amplitude
at $n_{c}$ is zero, the wave does not `see' the specific nonlinear value there. 
In other words, for all purposes, the nonlinearity at the void could be exactly $\chi$ and as a result, the situation is as in the usual homogeneous case, where there is a twisted mode. This observation also implies that, in a more general nonlinear impurity problem where there is a single site endowed with a nonlinearity value different from that of the background, there will be a twisted mode. 
In Figs.~\ref{fig1}(e) - (h) are sketched other modes, most of them unstable. We note these modes look like slightly deformed versions of modes (a) - (d), and they are all unstable. 

Figure 2 shows the perturbation eigenvalues for a chain of $N=100$ sites, for modes (a) - (d) shown in Fig.1 (localized modes) and modes (a) - (d) shown in Fig.4 (extended modes). For the localized modes, we can see that the symmetric mode around the void (shown in the inset of Fig.1) and the extended mode (f) in Fig.4, possess ordinary instabilities, while all the rest possess oscillatory instabilities.
\begin{figure}[t]
  \includegraphics[scale=0.5]{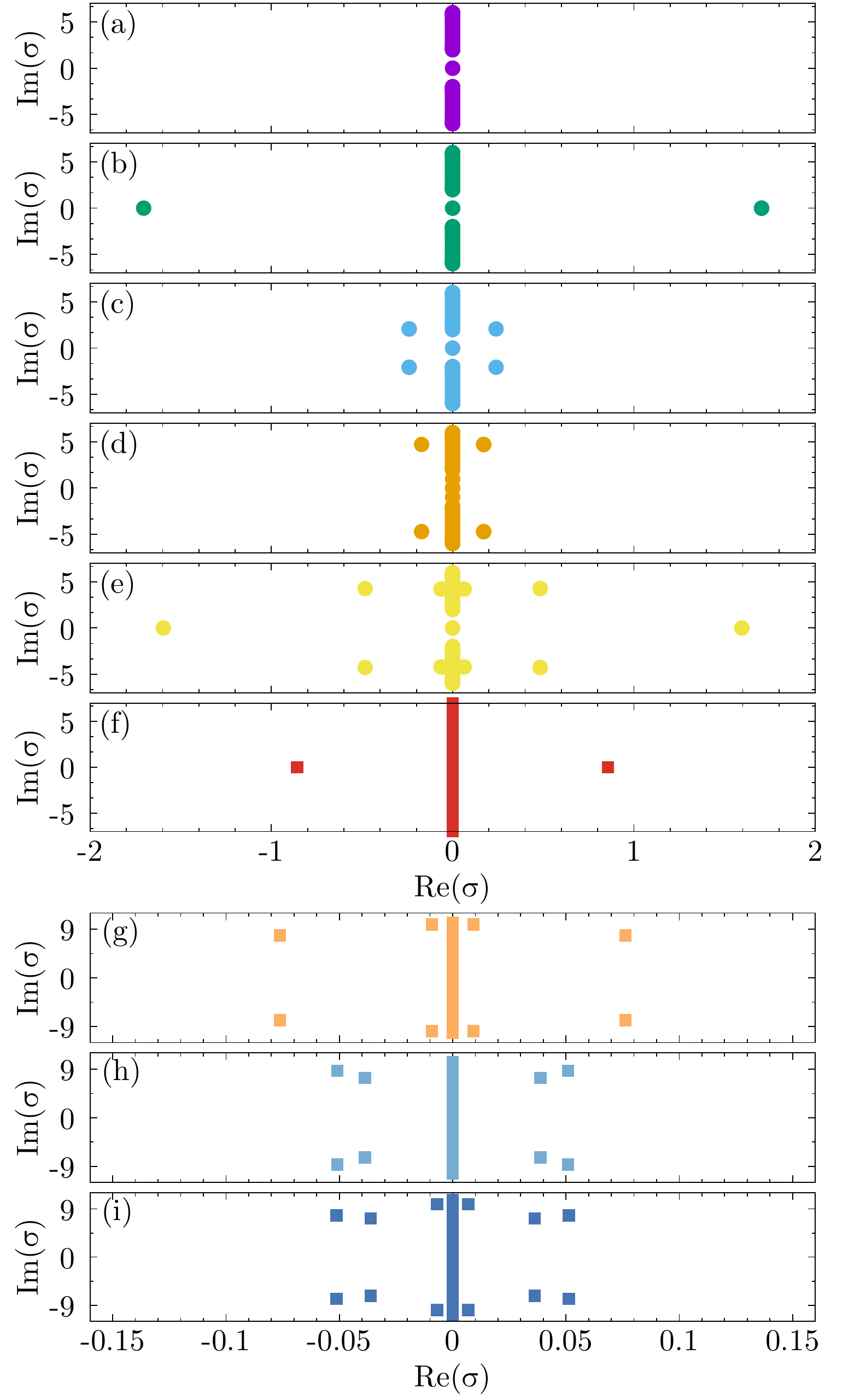}
  \caption{Perturbation eigenvalues for the modes shown in Fig. \ref{fig1} and Fig. \ref{fig3}. Boxes (a)-(e) correspond to modes (b)-(f) in Fig. 1, while boxes (f) - (i) corresponds to modes (a) - (d) in Fig. 4.}
  \label{newfigure2}
\end{figure}
Another thing we notice is that all the families show unstable behavior as the
eigenvalue approaches the linear band ($\lambda=2 V$). Since the modes with
eigenvalues close to the band are usually extended, one could infer that all
extended nonlinear modes will have a tendency towards instability. This behavior
will be confirmed in the next section.

\section{Extended modes}
Here we examine the existence and stability of solutions to Eq. (\ref{eq:2}) that
are extended in space. The idea is to make a comparison with the nonlinear
periodic case whose extended solution has the form
\begin{equation*} E_{n}(z)= A \exp[i (k n-\lambda z)], \end{equation*}
where $\lambda=2 V \cos(k)+\chi A^2$. As is well known, this solution possesses a
modulational instability for $0<k<\pi/2$ for $\chi \cos(k) > 0$~\cite{kivshar}.
This mechanism provides the seed for the formation of soliton modes. In our
case, the presence of the void impurity breaks the shift translational invariance making the single plane wane solution no longer a stationary state. What we expect, however, is that the presence
of the void will distort the plane wave modes, but will keep them extended all
over the lattice. It is natural then, to use the old plane waves, as seeds for
the iterative algorithm, in order to obtain the new stationary extended modes. 

\begin{figure}[t]
  \includegraphics[width=1.03\linewidth]{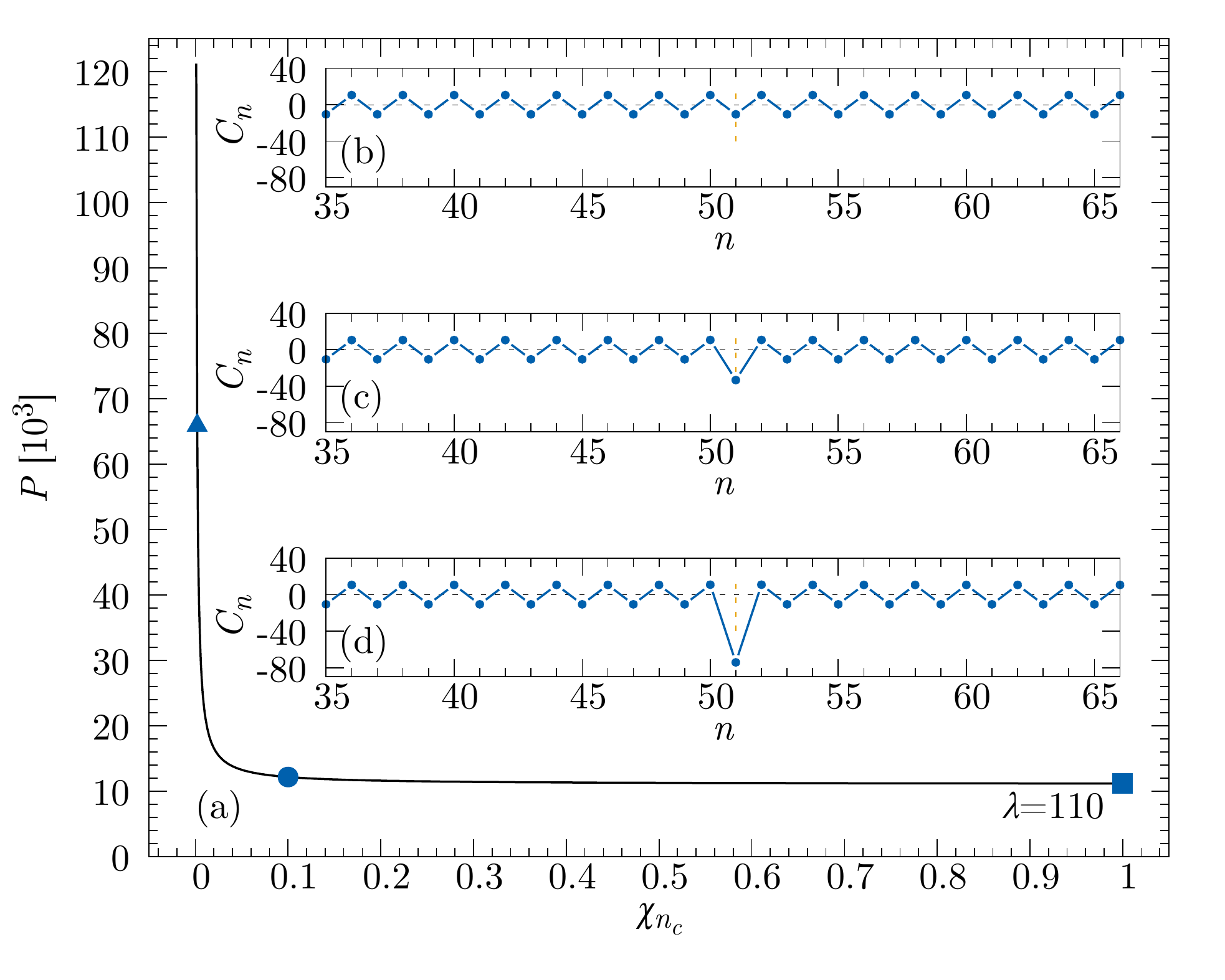}
  \caption{Power vs nonlinearity value at the void for $\lambda=110$.   The black line curve
  denotes stable solutions. Blue square, blue circle and blue triangle  correspond with amplitude
  profiles sketched in (b), (c) and (d), respectively.}
  \label{fig2}
\end{figure}

We begin our analysis by looking for nonlinear modes for the homogeneous case:
$\chi_n=1$ for all $n$. Here we find nonlinear plane waves as those predicted in
Ref.~\cite{kivshar}.  In order to understand how the void modifies the spatial
structure of extended solutions, we begin analyzing how they become altered when
we break the nonlinearity homogeneity in the lattice. We will focus on the
staggered case ($k=\pi$) because this is the only value for $k$, aside from
$k=0$, where our iterative scheme converges to stationary solutions\cite{comment}. By setting
$\lambda/V=110$, we proceed to slightly perturb the $\chi_{n_c}$ value,
diminishing it steadily.  Under this procedure, plane waves transform into more
complex structures, where the modulus of the amplitude at the void site grows as
we move towards $\chi_{n_c}=0$ and $P$ also increases as can be seen from
Fig.~\ref{fig2}. This behavior is maintained and, as we approach $\chi_{n_c}=0$,
the optical power becomes huge, signaling that this kind of solution has a
singularity at $\chi_{n_c}=0$.  In order to compare, we plot the amplitude
profile for three modes belonging to this family: the homogeneous case,
Fig.~\ref{fig2}(b), marked with the blue square on the $P$ vs $\chi_{n_c}/V$
curve, and the perturbed cases, Fig.~\ref{fig2}(c) marked with the blue circle
on the same curve and Fig.~\ref{fig2}(d), marked with the blue triangle.  It is worth mentioning that this kind of solutions is 
linearly stable in their entire domain. We note that the curves look nearly periodic and staggered, which can be obtained if
$\chi |C_{n}|^2 \sim \chi A^2$ is close to $\chi_{n_{c}} |C_{n_{c}}|^2$, because in that case we are back to a nonlinear periodic lattice. Thus, from $\chi A^2=\chi_{n_{c}} |C_{n_{c}}|^2$ we have $|C_{n_{c}}|^2=(\chi/\chi_{n_{c}}) A^2$, implying the hyperbolic dependence of $|C_{n_{c}}|^2$ (and $P$) on $\chi_{n_{c}}$ observed in Fig.~\ref{fig2}(a).

\begin{figure}[htb!]
  \includegraphics[width=\linewidth]{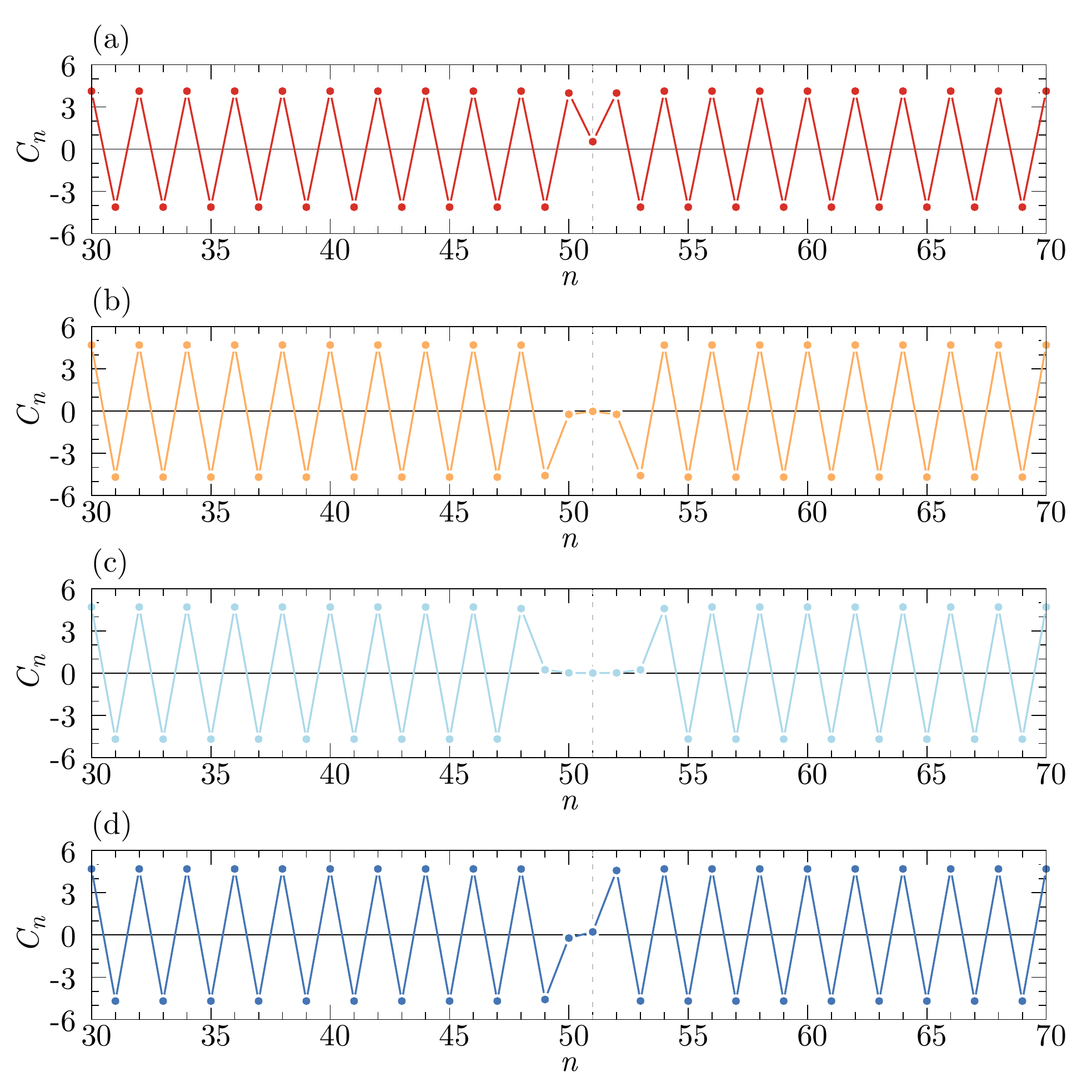}
  \caption{Amplitude profiles for discrete dark solitons in nonlinear optical
    lattices with a void. The dip is centered on the void's position. The modes
    sketched correspond to the 1-dip (a), the 3-dip (b), the 5-dip (c) and the
    2-dip (d), respectively. Vertical orange dashed line point out the location
    of void.}
  \label{fig3}
\end{figure}

\begin{figure}[t]
  \includegraphics[width=\linewidth]{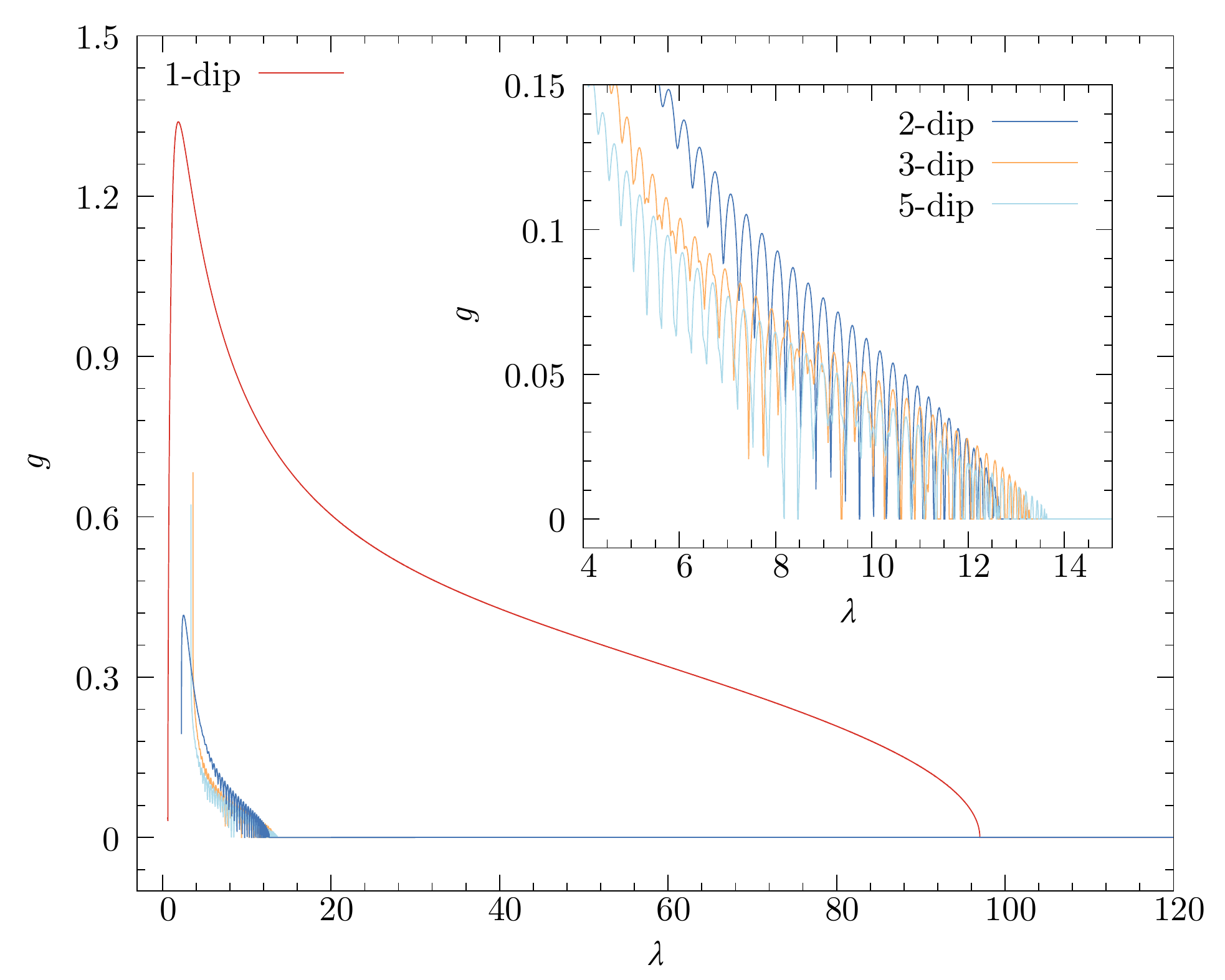}
  \caption{Gain vs eigenvalue stability diagram for dip type stationary
    solutions in nonlinear lattices with a void impurity. For 1-dip discrete
    dark solitons (red curve) the diagram shows that they are stable for
    $\lambda\gtrsim 97$. The 3-dip, 5-dip and 2-dip modes  are represented by
    blue, orange and cyan curves, respectively. The inset shows a zoom of
    their stability domain at low eigenvalue, showing that the 2-dip discrete
    dark solitons possesses the widest stable region, starting from values above
    $\lambda/V\sim 12$. }
  \label{fig4n}
\end{figure}

In spite of the asymptotic behavior for this family, once $\chi_{n_c}=0$ is
reached, in our iterative method, a new kind of solution appears. This solution
looks as a discrete dark soliton, i.e., a dip appears at $n_c$ into the plane
wave, as can be seen from Fig.~\ref{fig3}(a). However, these modes are unstable
in a wide range of $\lambda$'s, more specifically for $-2<\lambda/V<97$, i.e.,
this kind of solution only could be observed at high values of optical power.
Hereafter, we will refer to this specific solution as the `1-dip discrete dark
soliton', in order to compare it with subsequent stationary solutions that will
be discussed next. Given the simplicity of the spatial structure of the 1-dip
soliton, one could assume this solution to be the `fundamental' mode of the
system.  However, and as we pointed above, its stability is restricted to high
values of optical power only, so we will dispense with this term for this mode. Next, we search for other
dip-like modes with a wider stability range. We 
modify the 1-dip
solution by widening the dip of our initial seed of our iterative scheme finding
new solutions in the form of a `2-dip', `3 dip', and '5-dip' dark solitons. 
In Figure~\ref{fig3} we present the amplitude profiles for this new kind of dip
structures. 
They exhibit stability regions at low power which are wider compared
to the 1-dip discrete dark soliton.  Figure~\ref{fig3} also displays three new
discrete dark soliton structures, which exhibit a less trivial composition in
their dips. We note that the profile for the 3-dip and 5-dip discrete dark
solitons are even with respect to the position of the void ($n=n_{n_c}$). For
the mode sketched at Fig.~\ref{fig3}(d), the 2-dip discrete dark soliton, the
spatial symmetry is with respect to the midpoint between the $n_{c-1}$ and $n_{c}$
sites. The perturbation stability eigenvalues of these modes are shown on Fig. \ref{newfigure2}.
As we can see, the 1-dip solution possesses simple instability, while the rest display
oscillatory instabilities. Figure~\ref{fig4n} displays the linear stability of these dip solutions. It can
be quantified by the value of the instability gain $g$ of the mode, computed by
well-known linear stability analysis~\cite{stability}.  The quantity plotted is
the instability gain $g$ versus the mode eigenvalue $\lambda$. As we mentioned
at the beginning of this section, the 1-dip solution (red curve) is stable only
at  high $\lambda$ values.  For the other dip modes, the stability is different.
These stationary solutions persist in a stable regime for power values much
smaller that in the 1-dip case. The blue, orange and cyan curves correspond to
the 2-dip, 3-dip and 5-dip, respectively. Once they reach their instability
threshold, the gain parameter becomes positive and increases, but exhibiting a
complex oscillatory behavior. The instability windows for these dip-like modes
overlap at small $\lambda$, which hinders visualization. Therefore, we plot a
zoom of this zone in the top right inset of Fig.~\ref{fig4n}. 
These rapid oscillations of the gain near the instability thresholds are due to finite-size effects that cause repeated collisions of the unstable mode with the phonon band. As the size of the lattice increases, these oscillations decrease and cease altogether in the limit of an infinite lattice \cite{johansson}. 
We observe that the 2-dip solution becomes unstable for $\lambda/V\lesssim 12.6$,
with this value being the smallest one for the whole set of extended solutions
analyzed here. 
\section{Pulse transmission}
We consider now the transmission of a nonlinear pulse across the void impurity.
A possible choice for the shape of the initial pulse is the mode profile
corresponding to a homogeneous chain, computed  from Eq.
  (\ref{eq:2}) in the
absence of the void. We endow this pulse with a transversal momentum $k$ by
multiplying it by the phase $\exp(i k n)$. The mobile pulse thus created is not
really a solution of Eq. (\ref{eq:1}), but it is capable of traveling for long distances across the lattice with little radiation, when its width is much greater than the lattice spacing.

Another choice, one that does not require computation of the  stationary mode profile beforehand, is to take as an initial profile the discretized form of the soliton
solution of the continuous nonlinear Schr\"{o}dinger equation: $C_{n}(0)= A\
\mbox{sech}((A/\sqrt{2})(n-n_{0})) \exp(i k (n-n_{0}))$, where $n_{0}$ is the
position of the soliton center, $A$ is the height and $k$ the wave vector. As is
well known, this pulse will propagate for some distance before getting trapped
at some waveguide~\cite{vicencio,krolik}. Its effective mobility depends on the
height (width) of the pulse. 

\begin{figure}[t!]
  \includegraphics[width=\linewidth]{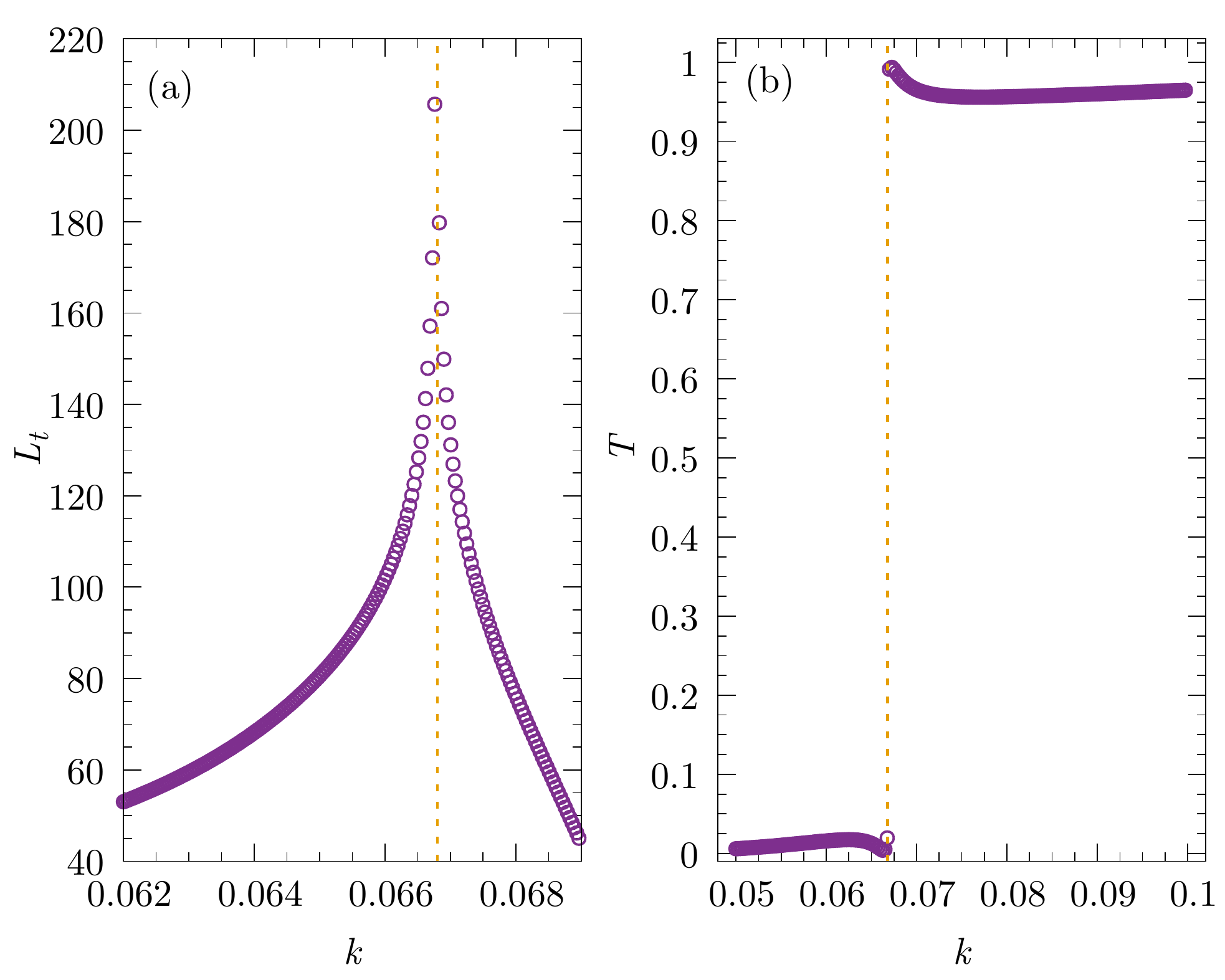}
  \caption{ Transmission of a nonlinear pulse across the void impurity.  (a)
  Extent of the trapping length $L_{t}$ (`time') as a function of the initial kick $k$.
  Absolute trapping occurs at $k_{c}=0.668$. (b) Pulse transmission coefficient $T$
  as a function of the initial kick $k$.}
  \label{newfig5}
\end{figure}

For definiteness, let us follow the first approach and consider an initial pulse
in the form of the discrete pulse solution of the homogeneous nonlinear chain, kicked
with a momentum $k$ pointing to the defect, and wide enough for good mobility.
When the pulse arrives at the void impurity position, it can get reflected,
transmitted and trapped. We define the transmitted (T), reflected (R) and trapped fraction (C) as
\be
T = {\sum_{n=n_{c}+\Delta}^{N} |C_{n}|^2\over{\sum_{n=1}^{N} |C_{n}|^2}}\hspace{0.5cm}
R = {\sum_{n=1}^{n_{c}-\Delta} |C_{n}|^2\over{\sum_{n=1}^{N} |C_{n}|^2}}\label{transmission}
\ee
and $C=\sum_{n_{c}-\Delta}^{n_{c}+\Delta}|C_{n}|^2/{\sum_{n=1}^{N} |C_{n}|^2}$, where $\Delta$ is 
the spatial extent of the trapping portion. Of course, we have $1=T+R+C$.

We proceed by sending pulses of different momenta $k$ and measuring $T$, $R$ and
$C$. A long array is used ($N=800$) as well as a proper longitudinal distance
$z$ (`time'), to permit a well-defined transmitted and reflected fraction.
Typically, the void impurity traps a portion of the wave and releases it some
time' later. We define the trapping length as the longitudinal distance (`time') over which the pulse remains trapped at the void position.
 
\begin{figure}[t]
  \includegraphics[width=\linewidth]{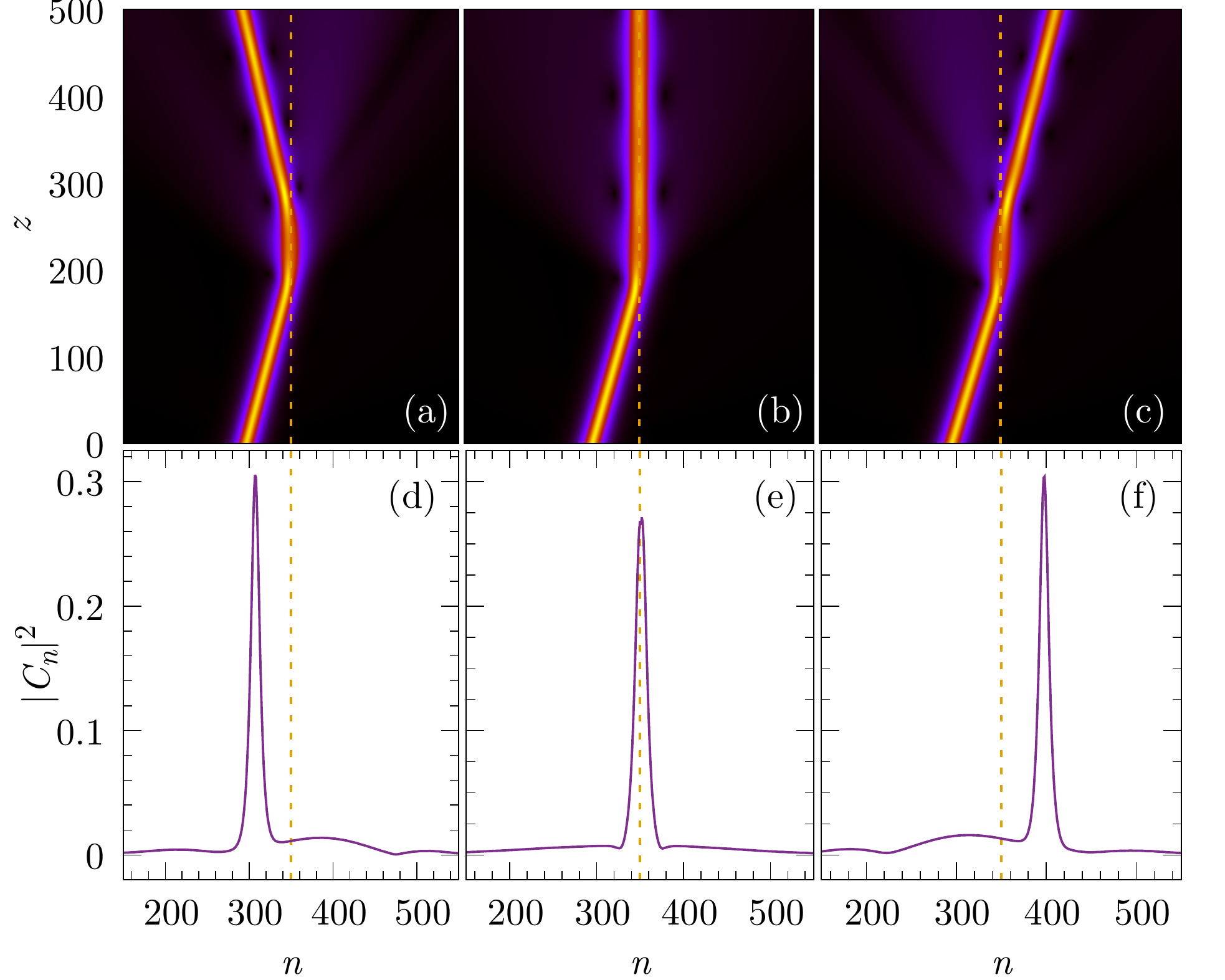}
  \caption{Top: propagation of the nonlinear pulse near the void impurity, for
    kick values near the critical one, $k_{c} = 0.0668$. Bottom: Output profiles
    near the critical $k_{c}$. Dashed line marks the position of the void
    impurity.} 
    \label{newfig6}
\end{figure}

During the collision process we observed the emission of some
radiation, as well as a critical $k$ value, $k_{c}=0.668$ at which the trapping length is `infinite', i.e., the void
impurity traps the incoming pulse `indefinitely'.  For kicks $k < k_{c}$ the wave
packet is reflected almost perfectly.  The transmittance coefficient is found
to be smaller that $0.02$, i.e., only up to 2\% of the wave energy is
transmitted. Near $k_{c}$ the amount of transmitted energy increase up to a
maximum of $2\%$, then it gets drastically reduced to approximately $1\%$ when
the beam is trapped by the defect. At $k_{c}$, the percentage of non-trapped
energy is around $2\%$, and it looks symmetrically distributed between reflected
and transmitted energy. Finally, when the kick is greater than $k_{c}$, the
situation is reversed, and most of the energy is transmitted. In all cases,
the transmitted and reflected portions propagate in a soliton-like manner.
Figure \ref{newfig5}(a) shows the trapping distance ('time') of the pulse as a function of the
transverse kick, while Fig. \ref{newfig5}(b) shows the transmission coefficient as a function
of the kick. As can be seen, there is an abrupt change in transmission at
$k=k_{c}$. Figure \ref{newfig6} shows the evolution of the pulse near the critical trapping
kick, and also snapshots of the spatial profile near the critical   kick. The vertical axis in Fig.\ref{newfig6} (top) corresponds to longitudinal propagation distance. All this phenomenology depends on the width of the pulse: For wide pulses the dynamical Peirls-Nabarro barrier is low and the nonlinear pulse can propagate for long distances inside the nonlinear chain. In this regime, the results obtained above apply. When is pulse is narrow, selftrapping effects will reduce its mobility, to the point that the incoming pulse might not even get to the scattering region, no matter how large the initial momentum. Trying to move a narrow pulse by simply applying a large momentum might result in the fragmentation of the pulse.

\section{Transmission of extended modes}
Let us consider the transmission of plane waves across the nonlinear ``void'' impurity. Unlike a linear system where the incident, reflected and transmitted fluxes can be identified clearly, in our case the nonlinearity couples the incoming and reflected fluxes, making it harder to identify them.  A simpler estimate of the effect of the void can be obtained by embedding the nonlinear segment in a linear lattice and computing the transmission across the segment with and without the void. This
procedure is physically justified since in real physical systems all nonlinear segments are of finite extent. It also allows for an unambiguous identification of the incident and reflected fluxes. We should also mention here the interesting recent development of non-reciprocal transmissions, where an acoustic or electromagnetic wave is made to propagate differently along two opposite directions. In that context, the method of embedding the nonlinear system in a linear lattice has proven very useful \cite{lepri}. 

Unlike the case of the transmission of finite pulses Eq. (\ref{transmission}), where the incoming, reflected and transmitting portions could be indentified with no ambiguities, now it becomes convenient to use an empirical procedure based on the recurrence equations Eqs.(\ref{eq:2}).
Since the system is nonlinear, it is convenient to use a backward iteration
scheme~\cite{delyon}: For a given wavevector $k$ and nonlinearity $\chi$, we
start to the right of $n=n_{c}$ where $C_{n} = T \exp(i k n)$ and iterate
Eq. (\ref{eq:2}) to the left going through the defect and further, until reaching the beginning of the segment. Then one can identify the transmission amplitude from the values of the amplitudes at the beginning of the nonlinear chain.

\begin{figure}[t]
  \includegraphics[width=\linewidth]{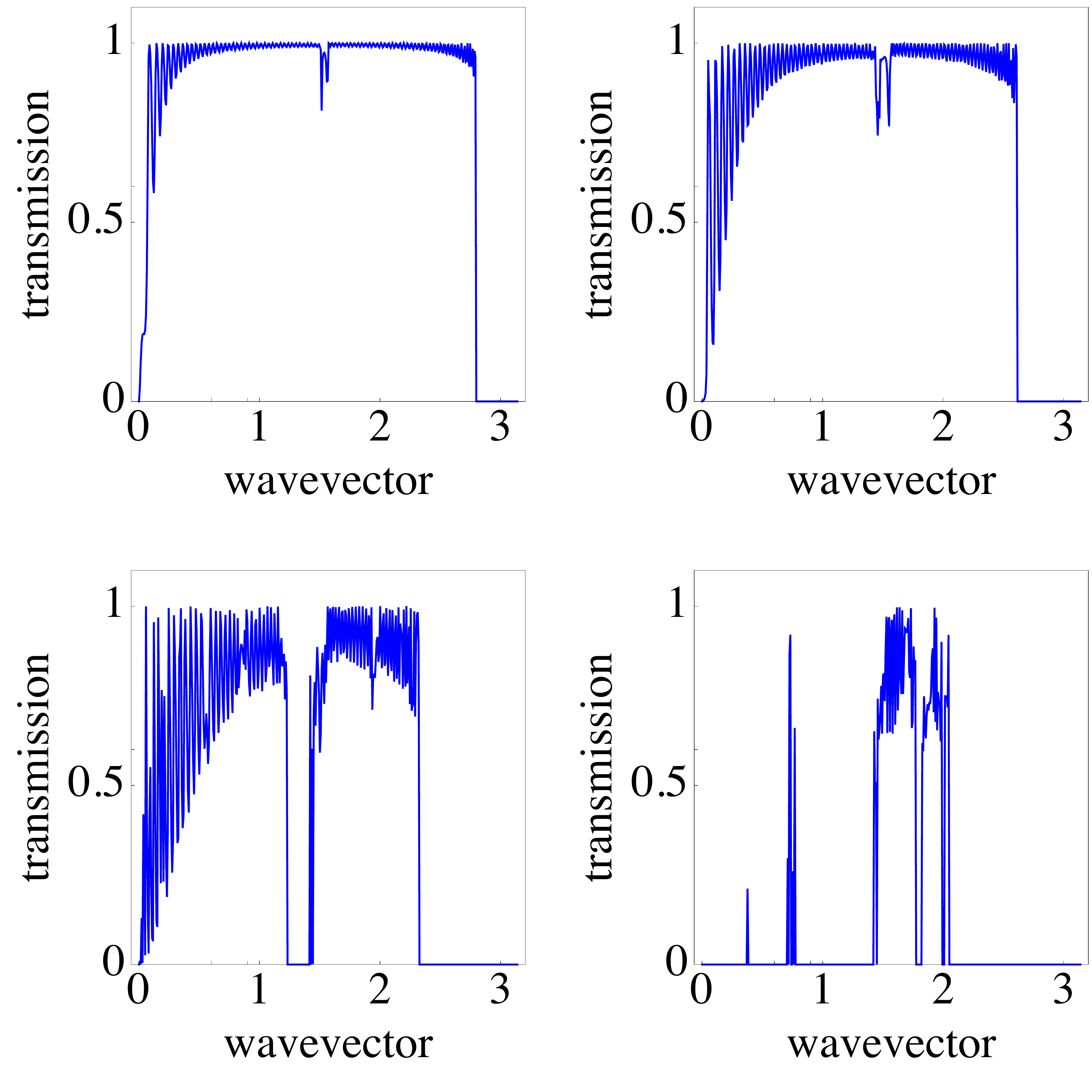}
  \caption{ Transmission coefficient of nonlinear plane waves versus wavevector,
    for several nonlinearity parameter values. Top left: $\chi=0.1$. Top right: $\chi=0.2$. Bottom left: $\chi=0.4$. Bottom right: $\chi=0.6$.}
  \label{fig8}
\end{figure}

Results from this procedure are shown in Fig.\ref{fig8}.  The transmittance
displays a rather complex structure as a function of $k$, with the creation of
more and more transmission gaps as nonlinearity increases. As nonlinearity is
increased, the overall transmittance decreases and disappears eventually near
$\chi=0.75$. Also, a computation of the output intensity as a function of the
input intensity (not shown) demonstrates that, for a given intensity, two or more
output intensities are possible, thus evidencing multistable behavior. This is
reminiscent of the behavior found for the homogeneous nonlinear
chain~\cite{delyon}.

\begin{figure}[t]
  \includegraphics[scale=0.13]{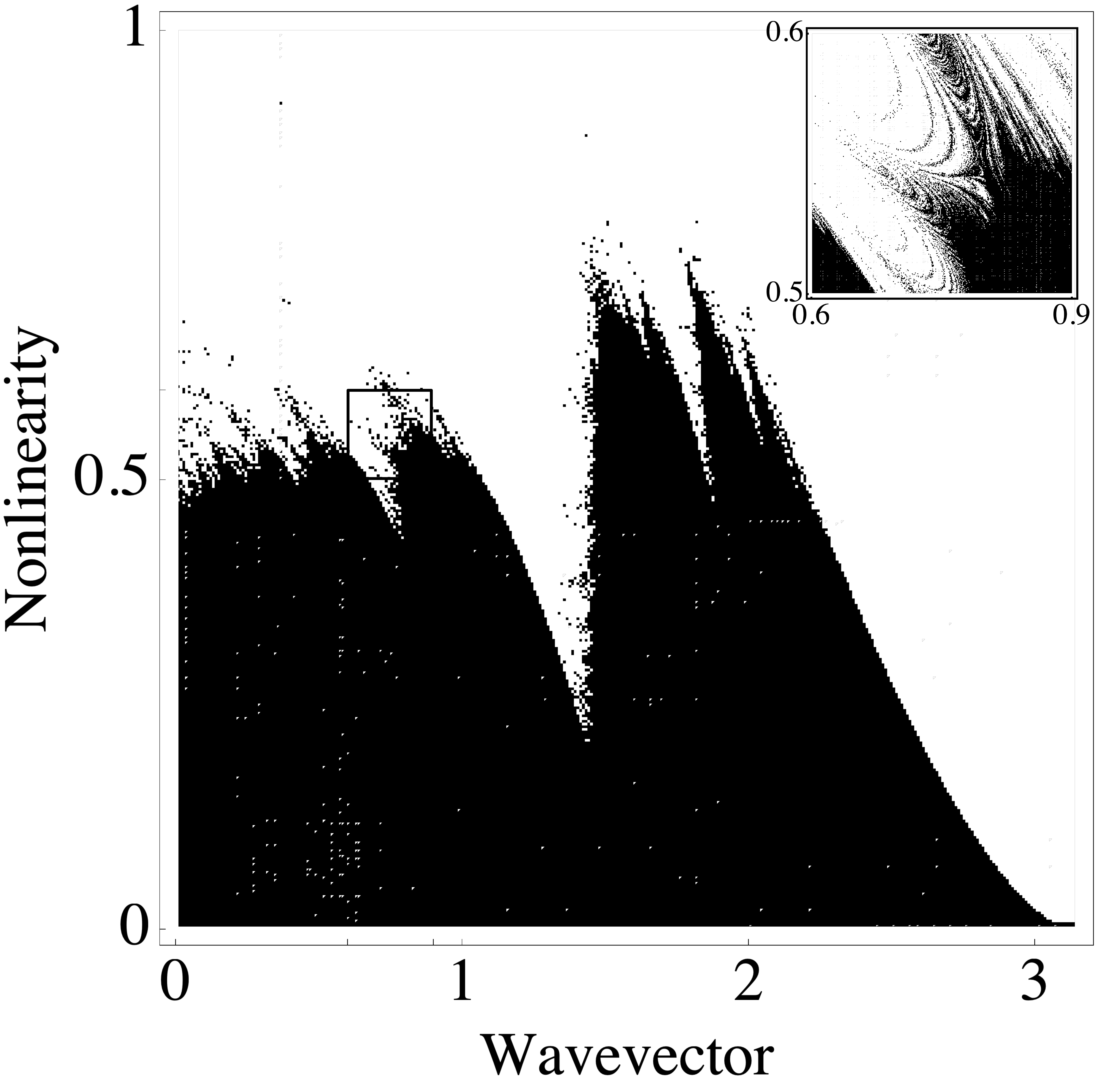}
  \caption{Transmittance diagram for nonlinear array with a void impurity.
  Dark (clear) regions denote transmitting(non-transmitting) regimes. Inset shows the transmission inside the parameter 
  region bounded by the square.}
  \label{fig9}
\end{figure}

\begin{figure}[t]
  \includegraphics[scale=0.25]{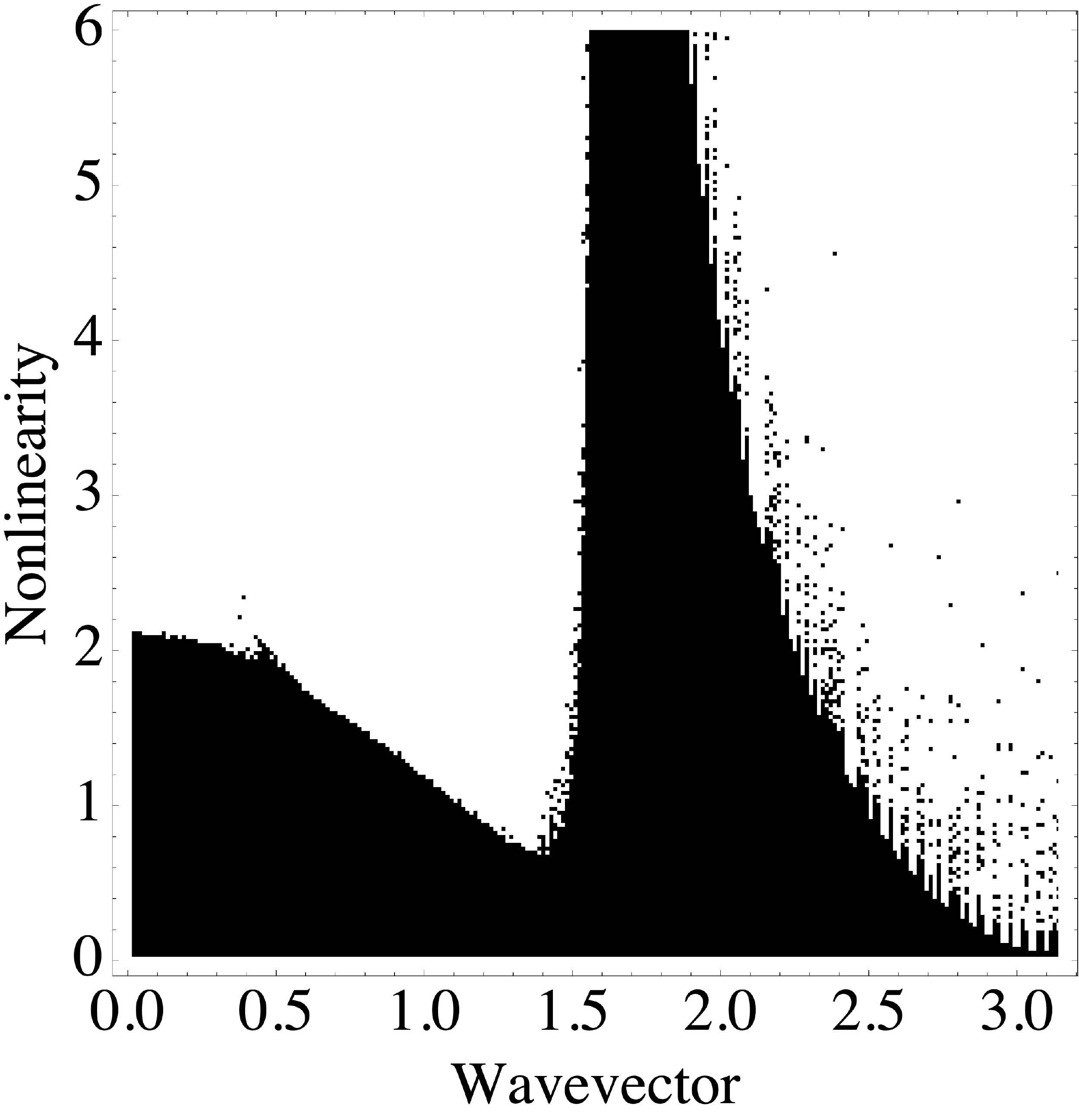}
  \caption{Transmittance diagram without the void impurity.
   Dark (clear) regions denote transmitting
    (non-transmitting) regimes. Notice the different nonlinearity scale.}
  \label{fig10}
\end{figure}

Figure \ref{fig9} shows a panoramic view of the transmittance situation by means
of a transmittance phase diagram. In this diagram, for a given $k$ and $\chi$,
we compute the transmission $T$, and if $T$ is greater than an arbitrarily
preset value $T_{c}$, we associate a value of ``1'' (passing) and mark a black
dot on the $k$-$\chi$ diagram; otherwise if $T<T_{c}$ we associate a value of
``0'' (non-passing) and make no marking. In this manner we obtain the
``passing/non-passing'' diagram shown in Fig.\ref{fig9} (we have taken $T_{c}=10^{-10}$.
Other values give similar results). Roughly speaking, $T$ decreases with an
increase in $\chi$ and, after a critical $\chi$ value, $T=0$ for any $k$. The
boundaries separating passing from non-passing regions are complex and
fractal-like. The complexity exhibited in this diagram, explains the main
features of Fig. \ref{fig8}, whose plots can be considered to be (roughly) cross sections
of Fig. \ref{fig9}. The irregularity of the boundaries of the passing region suggest that they could be fractal. We have computed the transmittance for a small region near the boundaries of the passing region in Fig. \ref{fig9} (see inset). The results show the persistence of irregularity to smaller scales. A quick box-counting computation places the fractal dimension at around $1.84$. To compare with the case of no impurity, we have also shown $T$ for
the periodic case with no void, in Fig. \ref{fig10}. Clearly, in the absence of the void 
the transmitting region is much larger.


\section{discussion}
In this work we have studied nonlinear localized and extended modes of a
nonlinear waveguide array that contains a single `void' waveguide that is, a
purely linear waveguide. Despite the simplicity of the model, the system
exhibits a rich behavior evidenced in the novel families of localized modes and
extended waves, and the interesting behavior of the transmission of pulses and
extended modes across the nonlinear void. The low-lying localized modes in the
vicinity of the void display a great deal of instability, specially at low power
levels, when the mode spatial extent is the greatest. The simplest mode is
doubly-degenerate, with its center lying one site to the left or one site to the
right of the void position. This is in marked contrast with the case of a linear
or nonlinear impurity in a linear chain where the mode center is located  at the
impurity position. Next, we investigated the extended modes of the array, modes
which in the absence of the void are nonlinear plane waves $A \exp(i (k n
-\lambda z))$ with $\lambda=2 V \cos(k)+\chi A^2$. In the presence of the void
impurity these modes constitute natural starting points for the Newton-Raphson
iterative method. In that way, we have found extended modes that are different
in character to the usual plane wave. Instead of a continuous deformation of the
wave in the vicinity of the void, these solutions display a rather abrupt
profile change. Another interesting feature is that only plane wave modes with
$k=0$ or $k=\pi$ lead to extended modes\cite{comment}. Of these two, only the seed with
$k=\pi$ lead to a solution with a stability region. It should be mentioned here that the existence of stable dark modes with $k=\pi$ in the focussing regime implies--via the staggered-unstaggered transformation ($\lambda\rightarrow -\lambda, \chi\rightarrow -\chi, C_{n}\rightarrow (-1)^n C_{n}$ ), the existence of a defocusing system with stable dark states with $k=0$.

In order to ascertain the
role of the nonlinearity jump at the void, we proceeded in a continuous manner
decreasing the value of the nonlinearity at the void position from $\chi=1$ down
to $\chi=0$, and examining the extended mode produced. We observed that at
$\chi\rightarrow 0$ a singularity is approached, beyond which the system enters
a new phase where the extended modes have the form of a constant background with
one or several dips at the position of the void.  The analysis of the stability
of these modes reveals that the mode with 2 dips is the one with a larger
stability window. Next, we
investigated the transmission of a nonlinear pulse across the void impurity. By
using a wide nonlinear pulse, based on the discrete soliton solution of the homogeneous nonlinear chain, we were
able to find well-defined transmittance, reflectance and trapping at the
impurity. The different regimes are separated by a critical wavevector, and
characterized by an abrupt change in the transmission. Finally, we examined
numerically the transmission of a nonlinear plane wave across the void. The
transmission shows a complex structure as a function of wavevector, with the
creation of more and more transmission gaps as the nonlinearity of the
background is increased. In order to obtain a general bird's eye view of the
transmittance properties we defined an overall transmittance by passing and
non-passing regimes. In nonlinearity-wavevector parameter space, the overall
transmittance decreases as a function of nonlinearity, and disappears
eventually. One interesting feature is that the boundary separating passing from
non-passing regimes is complex and fractal-like.
 
We anticipate that our results will help pave the way for experimental
observations of these novel type of nonlinear optical modes in photonic
lattices.

\acknowledgments
The authors are grateful to the anonymous referees for comments and suggestions that helped increase significantly this article. M.I.M acknowledges support from Fondecyt Grant 1160177. A.A.S. acknowledges support by the Australian Research Council (DP160100619 and DP190100277). The authors are grateful to P. G. Kevrekidis for useful discussions.


\begin{thebibliography}{99}
\bibitem{comment} This stems from the continuity equation $\partial \rho/\partial t + \mbox{div} J=0$, where $\rho$ is the density and $J$ is the current. For a periodic, discrete system $\rho=|E_{n}(z)|^2$ and
 
$J_{n} = 2 V \mbox{Im}[ E_{n}^{*}(z) E_{n-1}(z) ]$. If we use a nonlinear plane wave as an ansatz to find an extended mode for the impurity system, $E_{n}(z)=A \exp[i (k n-\lambda z)]$ with $\lambda=2 V \cos(k)+\chi|A|^2$, which implies $\rho=|A|^2$ and $\partial \rho/\partial z =0$, while $\mbox{div} J = J_{n+1}-J_{n}= 2 V |A|^2 \mbox{Im}[ \exp(i k)-\exp(-i k) ]\sim \sin(k)$. Thus, the continuity equation is fulfilled at $k=0,\pm \pi$ only.
\end{thebibliography}
\end{document}